\documentclass[prl,twocolumn]{revtex4}

\usepackage{graphicx}
\usepackage{dcolumn}
\usepackage{amsmath}
\usepackage {epsfig}

\def\lbets{$\lambda-(BETS)_2FeCl_4$~}
\def\lbetsgabr{$\lambda-(BETS)_2Ga_{1-x}Fe_xBr_yCl_{4-y}$~}
\def\pid{$\pi-d$~}

\makeatother
\begin{document}
\title[Short Title]{Magneto-optical probe of  antiferromagnetic sub-phases in a
\pid electronic system}
\author{I.Rutel$^1$, S. Okubo$^2$, H. Ohta$^2$, H. Tanaka$^3$, H. Kobayashi$^3$,J. Brooks$^1$ }
\affiliation{$^1$National High Magnetic Field Laboratory and
Physics Department, Florida State University, Tallahassee, FL
32310 USA} \affiliation{$^2$Molecular Photoscience Research Center
and Venture Business Laboratory,
Kobe University, Kobe, Japan}
\affiliation{$^3$Institute for Molecular Science, Okazaki, Japan}
\date{\today}

\begin{abstract}
The highly insulating antiferromagnetic phase of the magnetic-
organic conductor \lbets has been probed by resonant mm wave
methods vs. magnetic field, temperature, and frequency. Our
results show evidence for magnetic sub-phases within the
temperature-field phase diagram as previously predicted by
Brossard \emph{et al.} [Eur. Phys. J. B 1, 439(1998)].
\end{abstract}

\maketitle

The discovery of magnetic field induced superconductivity in
\lbets \cite{uji} has drawn attention to the unusually large \pid
electron spin exchange mechanisms\cite{mori} in molecular systems
where magnetic order in the d electron system strongly influences
the behavior of the conducting $\pi$ electron system. The more
general \lbetsgabr class of organic conductors, with localized
magnetic moments at the anion sites, and conduction electrons in
the molecular-cation layers, exhibit competition between magnetic,
metallic, insulating, and superconducting ground
states\cite{tanaka}. The magnetic field dependent phase diagram of
the \lbets material is shown in Fig. 1. For H =0, and below the
N\'{e}el temperature\cite{neel} ($T_N = 8$ K), \lbets enters a
highly insulating antiferromagnetic (AF) phase\cite{bros}. A
spin-flop transition to a canted antiferromagnetic (CAF) phase
occurs near 1 T, and above 11 T, a paramagnetic metallic (PM)
phase appears. At higher magnetic fields (parallel to the the
conducting molecular planes), field induced superconductivity
(FISC) is stabilized below 5 K between 18 and 45 T\cite{bal}. The
FISC state involves the cancellation of the exchange field by the
external magnetic field. Although the CAF-PM transition is nearly
independent of field direction, the PM-FISC transition requires
careful alignment of the field in the \emph{a}-\emph{c} molecular
planes to avoid orbital dissipation in the superconducting phase.

    The main purpose of this work is to focus on the process by which the
magnetic field destroys the CAF state. This process has been
previously considered theoretically by Brossard \emph{et al.} who
use a generalized Kondo lattice model with two d-electron spin
sites and four $\pi$-electron HOMO bands\cite{bros}. The moments
involved can arise from the Fe$^{3+}$ and spin 1/2 conduction
electrons, which can interact through RKKY
interactions\cite{bros}. One key feature of their model was the
strong interdependence of the Fe$^{3+}$ spin configurations with
the electronic structure of the conduction electrons, which at the
two extremes yields for AF order - an insulator, and for PM order
- a finite Fermi surface. From this they predicted that increasing
magnetic field could lead to a sequence of different canting
angles as the PM phase was approached. The main result of our
work, which is in accord with their predictions, is that the CAF
spin structure appears to change in steps as the field (and/or
temperature) is increased, and that these discrete changes in spin
structure cause corresponding steps in the complex conductivity as
the PM phase is approached.

\begin{figure}[]
\epsfig{file=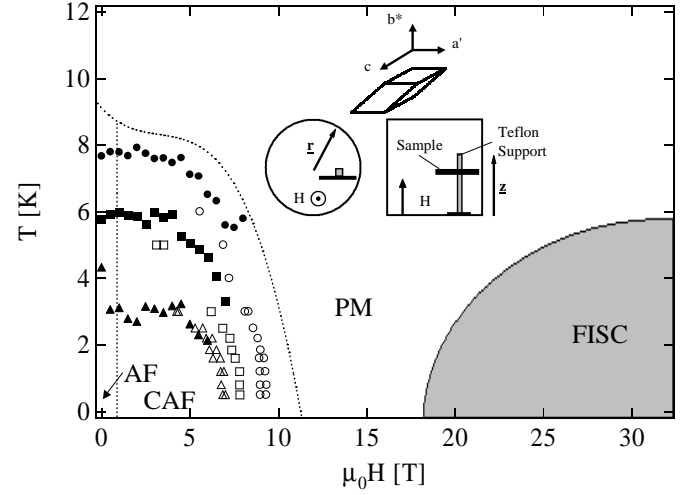,height=8.75cm,angle=-90} \caption{Phase diagram
of \lbets . The four phases identified are antiferromagnetic (AF),
canted antiferromagnetic (CAF), paramagnetic metal (PM), and field
induced superconducting  (FISC) states. Closed (open) symbols
represent the temperature (field) sweep measurements of the CAF
sub-phase structure: A-circles, B-squares, C-triangles (see text).
Schematic: Top figure: sample geometry and principle axes. Lower
figures: axial and side views of sample orientation. The sample
position shown is for the \emph{c}-axis$\parallel\hat{r}$ and the
\emph{c}-axis$\perp\hat{z}$, where $\hat{r}$ and $\hat{z}$ are the
unit directions for the axial and radial components of the
cylindrical cavity.}\label{fig1}
\end{figure}

    Since the resistance below $T_N$ rapidly becomes unmeasurable,
we have probed the antiferromagnetic phase through the use of ac
magnetoconductivty, and electron magnetic resonance in the 40 to
110 GHz frequency range. In previous X-band studies\cite{bros},
the temperature dependent electron spin (ESR) and
antiferromagnetic (AFR) resonance features (below $T_N$) have been
investigated vs. temperature and sample orientation with respect
to magnetic field. In the present study we extend the range of
investigation of the CAF state (and also the FISC state) to a
broader range of frequencies, temperatures, and magnetic fields. A
mm wave resonant cavity perturbation method is employed\cite{mvna}
where the sample orientation in the cavity can be changed with
respect to the field orientation. Samples were synthesized by
electrochemical methods\cite{kob1}.  The single crystal used in
this work, which was 0.97x0.10x0.10 $mm^3$ in size, was greased to
a small teflon mount inside the cylindrical cavity resonator. In
Fig. 1 the convention for orientation is given in terms of the
principle axes of the crystal, and the radial and axial ($\hat
z\parallel\vec H$ in all cases) unit vectors associated with the
cavity.  A control experiment insured that the grease and teflon
support did not contribute a significant background or spurious
ESR signal.

\begin{figure}[]
\epsfig{file=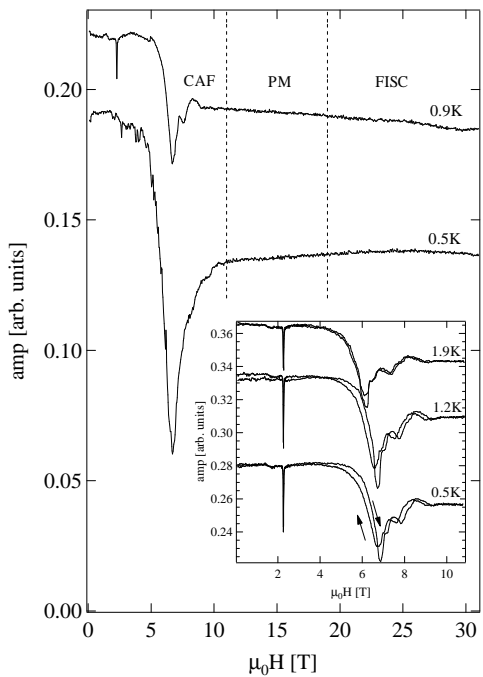,width=6.3cm} \caption{Low temperature ($T <
T_N$) field-dependent mm wave cavity response of \lbets for
$\emph{c}\parallel\hat z$, $\emph{c}\parallel\vec H$. (All data is
for $\nu$ = 66.9 GHz, except the 0.5 K data in the main panel
where $\nu$ = 75.4 GHz.) The sharp dip near 2 T is the ESR line,
and the features in the range 5 to 10 T arise from sub-phases in
the antiferromagnetic CAF state (see text). Although the sample
was aligned to allow the high field superconducting (FISC) state
to be stabilized, only weak features associated with a metal to
superconducting transition were detected. Inset: Full field sweeps
to 11 T. Hysteresis is observed in the sub-phase structure in the
CAF state.}\label{fig2}
\end{figure}

    Experiments were carried out in a 30 T resistive magnet and an 8 T
superconducting magnet where either temperature or field were held
fixed at a specific resonant cavity mode(frequency). A helium
3-probe was modified to allow temperatures down to 0.5 K without
interference from liquid dielectric effects. Our results are
presented in Figs. 2-5, and the phase diagram of the CAF state,
based on all of our results, is shown in Fig. 1.  Representative
results are shown for magnetic field sweeps in Fig. 2 for
\emph{c}$\parallel\hat z$ for several temperatures below T$_N$ up
to 30 T. The signal is the phase-locked amplitude of the cavity
response, where resonant frequency is allowed to change (via
feedback) to keep the phase reference in quadrature.
    In general, there are two signals in the mm wave cavity response, a
\underbar{resonant signal} (i.e. $h\nu = g\mu B$) due to electron
spin, and \underbar{non-resonant structure} where there is no
functional relationship between frequency and field position. The
latter arises from changes in the ac complex conductivity of the
material. For increasing field an ESR adsorption line appears,
followed by a change in background signal that is punctuated by
dips, peaks, and/or shoulders. The field positions of the CAF-PM
and PM-FISC phase boundaries observed in transport data are also
shown in Fig. 2 for comparison. The features at higher field are
not resonant, but reflect step-like changes in the ac
magnetoconductivity. As discussed below, we interpret these
features as sub-phase transitions in the CAF state for increasing
field. An exploration of the high field region where the field
induced superconducting phase is stabilized did not show any
systematic signature, other than slight, hysteretic loop-like
features that were not readily reproducible.

    We first address the magnetic resonance data, that is the
ESR and AFR measurements used to determine the g-factor and
spin-flop field respectively. A temperature dependent
investigation of the ESR line, shown in Fig. 3, reveals similar
features to those previously reported by Brossard \emph{et al.},
with the g-value exhibiting a non-monotonic temperature dependence
below $T_N$. The linear frequency-field dependence of the ESR line
yielded a g-value of 2.05 for sample orientations away from those
where the AFR signal was observable, as shown in Fig. 3b.

\begin{figure}[]
\epsfig{file=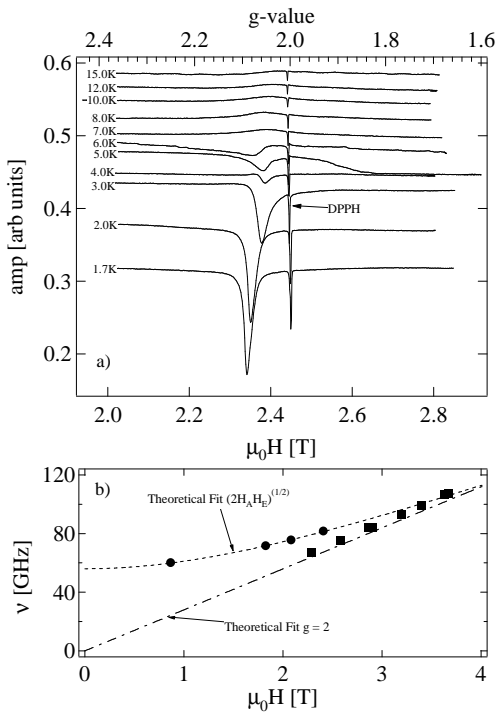,width=6.5cm} \caption{ a) Detailed temperature
dependence of the ESR line (with respect to the DPPH marker);
$\emph{a}\parallel\hat r$, $\emph{c}\parallel\vec H$ , $\nu$ =66.9
GHz. Below $T_N$ the g-factor is non-monotonic with decreasing
temperature. b) Solid circles - AFR vs. frequency for
$\emph{c}\parallel\hat r$, $\emph{a}\parallel\vec H$; Solid
squares - ESR vs. frequency for $\emph{c}\parallel\hat z$,
$\emph{c}\parallel\vec H$}\label{fig3}
\end{figure}

For the orientation \emph{a}$\parallel\vec
H$,\emph{c}$\parallel\hat r$, we measured the field position of
the the AFR signal over our accessible range of frequency, as
plotted in Fig. 3b (see also representative trace of data for
$\parallel\vec H$,\emph{c}$\parallel\hat r$ data in Fig. 5 below).
Although the frequency dependence of the AFR signal is
characteristic of the condition with the field along the hard
axis\cite{foner,hagiwara,coulon,torrance}, the precise field and
magnetic axis relationship is difficult to determine since the
easy axis lies about 35 degrees away from the c-axis, and the
system is, moreover, canted above the spin flop field.
Nevertheless, an extrapolation of the AFR resonance to the zero
field gives a gap frequency which corresponds (through $
h\nu/g\mu_B$), to a characteristic field of 1.7 T. This value,
which is related to the product of the exchange field and the
anisotropy field, is comparable to the spin-flop field (1.1 T)
reported in magnetization studies\cite{bros}.
\begin{figure}[]
\epsfig{file=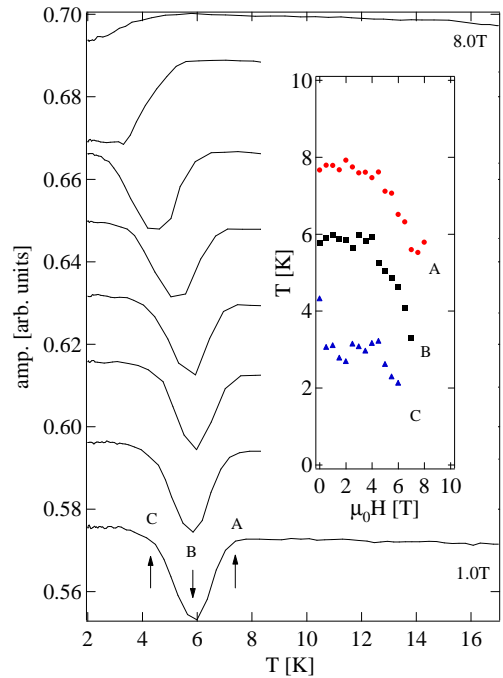,width=6.5cm}
 \caption{Temperature dependence of the cavity response at
 different fields
 for \lbets; $\emph{c}\parallel\vec H$, $\emph{a}\parallel\hat r$, $\nu$ = 64.927 GHz.
 Three changes
 in the cavity absorption with temperature are observed at points A, B, and C.
 Inset: phase boundaries of of A, B, and C with temperature and field.}\label{fig4}
\end{figure}
    We now turn to the non-resonant features in the ac magnetoconductivity
measurements, exhibited in Figs. 2, 4, and 5. We have identified
three recurrent features in the temperature and field dependent
cavity response below $T_N$ which appear regardless of frequency
or sample orientation, labelled as A, B, and C. When plotted vs. T
and H, these features are clearly a manifestation of the CAF phase
diagram. We interpret these features as changes in the
magnetoconductiviy of the sample due to changes in the electronic
structure in the $\pi$-electron system. This can cause changes in
the skin depth of the sample and therefore details of the response
of the resonant cavity response will change. We cannot, however,
at this stage assign these changes to specific values of the
complex conductivity of the material.

    Fig. 4 shows the temperature dependence of the cavity response for
fields between 1 and 8 T.  Below 8 K we identify three separate
points that indicate a change in the cavity response. As the CAF
phase is entered from higher temperatures, the signal is first
observed to fall (A), then to rise (B), and then to approach a
value after (C) that is larger than that observed above $T_N$. The
T-H plot of these features,  shown in the inset of Fig. 4, follow
the general shape of the CAF phase boundary, and indicate the
presence of sub-phases. To map out the low temperature sub-phase
boundaries we used field sweeps at constant temperature (Figs. 2
and 5). Here, similar structures were identified that allowed
extension of the sub-phase transition lines to lower temperatures,
as summarized in Fig. 1.

    An important point in our interpretation is that the field and
temperature position of the features (A,B,C) are non-resonant, and
only weakly dependent on sample orientation. Figure 5 serves to
demonstrate these assertions, and also sheds further light on the
nature of frequency dependent resonant cavity measurements. Here
the sample was measured for three different orientations in field,
different frequencies, and different temperatures. The ESR line is
prominent and absorptive at two of the frequencies, and for
$\emph{c}\parallel\hat r$, $\emph{a}\parallel\vec H$ the AFR
signal is observed. Strong ESR or AFR signals are good indications
of the sample's coupling to  $\vec H_{ac}$. (A detailed analysis
of cavity mode effects will be published elsewhere\cite{okubu}.)
In contrast, the ESR line is weak and inverted for the 71.27 GHz
data. Here also the field dependence of the background is
reversed. It is apparent that the overall signal depends on how
the sample is coupled with an $\vec E_{ac}$ or $\vec H_{ac}$
excitation mode, as indicated by the intensity of the ESR
features. However, the ESR line positions rigourously follow the
linear frequency dependence shown in Fig. 3b. Returning to the
non-resonant structures, the appearance of these features depends
on the cavity geometry and the resonant mode employed. The
dissipation of the cavity is governed by the surface resistance,
$R_s$, ($\propto Re{\sqrt{\frac{1}{\hat\sigma}}}$)\cite{hillsc}.
The response is a measure of the complex magneto-optical
conductivity of the material via the skin depth of the sample,
$\hat\sigma$. Although different shaped structures appear, their
assignments (A, B, or C) on the T-H phase diagram remain
congruous. Unlike the ESR lines, these features show no systematic
dependence on frequency.

    The premise that the spin configuration of the $Fe^{3+}$ ions
affects the electronic structure of the $\pi$-electron system has
two extremes: the highly insulating AF phase for H = 0, and the
paramagnetic-metallic phase\cite{ujifermi} for H $\gtrsim$ 11 T.
The energy difference between these two states, based on numerical
estimates from the Kondo lattice model\cite{bros}, is 0.5 meV, or
about 6 K, not far from $T_N$. Between these two extremes, our
data indicate that the electronic structure changes in a step-like
manner with either increasing temperature or magnetic field as the
CAF-PM state is approached. Since the d- and $\pi$-electron spin
configuration is a highly coupled system, it is an over
simplification to treat the spin projection $S_Z = \frac{5}{2}$
for the $Fe^{3+}$ ions as a proper eigenvalue in the relationship
${\cal H} = \frac{-e}{2m_{e}c}\mid\textbf{B}\mid S_{Z}$ where
$\textbf{B}$ is the magnetic field. Although heuristic, we may
replace $S_{Z}$ with an effective spin $S_{eff}$ to address the
step-wise changes observed in the CAF state. Using the
prescription that $k_BT_{(A,B,C)}(B=0)$ corresponds to
$\frac{-e}{2m_{e}c}\mid\textbf{B}_{(A,B,C)}\mid S_{eff}$, we
obtain from Fig. 1 that $S_{eff}$ = 0.29, 0.25, and 0.15 for
transitions A,B, and C respectively. This would indicate a
stepwise increase in the effective spin projection as the PM phase
is approached. Notably, $S_{eff}$ in the CAF phase is
significantly less than 5/2.
\begin{figure}[]
\epsfig{file=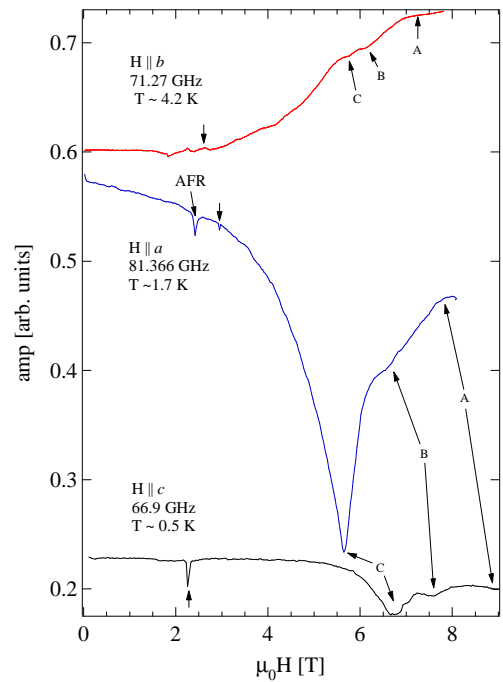,width=6.5cm} \caption{Magnetic field dependence
of cavity response for  \lbets in three different sample
orientations, frequencies, and temperatures. Arrows indicate the
position of the ESR lines, and the AFR line for
$\emph{a}\parallel\vec H$ is also indicated. The step-like
structures (A,B, C) between 5 and 10 T appear in all
cases.}\label{fig5}
\end{figure}

    In summary, we present new features in the phase diagram of the \pid electron system \lbets.
Using a high frequency resonant cavity measurement we show
sub-phase structure within the low temperature, low field CAF
state. These sub-phase transitions are the result of changes in
the spin-configuration in the CAF state which in turn cause
changes in the $\pi$-electronic structure. These changes are
therefore observable as changes in the complex conductivity of the
sample in a resonant mm wave cavity. Although stepwise, the
transitions are not strictly quantized, and the effective moment
is reduced from that of the free d-electron spin. These results
are consistent with a highly collective interaction between the
$\pi$-and d-electron systems.
\begin{acknowledgments}
We are grateful to  S. Wiegers, J. Rook, and  H. van Luong for
allowing us to use the new  Nijmegen-Amsterdam Magnet Laboratory
resistive magnet prior to formal commissioning the high field
measurements carried out it this work. The authors also
acknowledge Venture Business Laboratory, Kobe University. This
work is supported by NSF-DMR 99-71474. The mm wave facility was
supported through NHMFL/IHRP 5031.
\end{acknowledgments}


\begin{references}
\bibitem{neel} It is not clear at this point if a Peierls-like $\pi$ electron transition and
and an antiferromagnetic d-electron transition are, by
coincidence, at the same temperature. However, for convenience, in
this work we will refer to the transition as $T_N$. See also Ref.
\cite{tanaka}.
\end{references}
\end{document}